\documentclass[12pt]{article}
\usepackage{graphicx}
\usepackage{cite}
\usepackage{amsmath}

\begin{document}

\title{Measuring the Subjective Passage of Time}

\author{ Serge Galam\thanks{serge.galam@sciencespo.fr} \\
CEVIPOF - Centre for Political Research, Sciences Po and CNRS,\\
1, Place Saint Thomas d'Aquin, Paris 75007, France}

\date{January 8, 2024}

\maketitle

\begin{abstract}

I  address the question of measuring the subjective perception of the passage of time at the individual level in relation to its objective duration using a physicist-type treatment. A simple model is thus built in terms of a very small number of equations, which is found to yield a quantitative framework for individual subjective perception of time. Given an objective unit of time like the year, the hypothesis is to introduce a subjective unit of time for each person, which is inversely proportional to the number of objective units already experienced by that person. A power law is also considered. The model is then implemented by stacking individual clocks, each being linked to social rituals, which mark and shape the specific time of an individual throughout their life. For each person, the first of these events is their own birth. A future horizon, as well as an origin of the past, are then defined, along with a speed of subjective passage of time. The model shows that the price for the first ritualized socialization is to exit eternity in terms of a future to be lived with the simultaneous reward of experiencing a moment of infinity analogous to that of birth. The results recover common feelings about the passage of time over a lifetime. In particular, the fact that time passes more quickly with age.

\end{abstract}

\newpage
\section{Introduction}

Clocks are everywhere shaping everyone path of life at both individual and collective levels. Moreover, the measure of the objective time has reached an incredible precision with the Cesium atomic clocks, which serves as the world standard  \cite{ces}. 

The goal was to construct a time reference frame that is independent of individuals and their histories. The objective time is  measured using seconds, minutes, hours, weeks, months, years, centuries, and millennia. Those units of time are fixed and unaffected by the passage of time itself. 

Once defined, the interest of these units lies in their ability to be counted and added objectively, systematically, and, above all, understood by everyone. Such an objective requires, to be achievable, the definition of a counting origin. At this stage, while the choice of an origin is arbitrary, it requires to be acknowledged and used by everyone. In fact, from a practical point of view, the nature of the origin does not matters to the unification of a unique metric of time. Regarding years, it is often an event marking the beginning of a religion, such as the birth of Christ for the Western calendar, now used worldwide. 

Choosing and agreeing on unit specific of time allows identifying all events in relation to each other, in a linear and orderly manner. The identification of an event being the number of time units that have elapsed from the counting origin to the occurrence of the event. The notion of duration comes in naturally with the number of time units elapsed. Therefore, the duration defined is the same for everyone, whether an individual or a machine. And indeed, this was the intended purpose to bring order to the "disorder" of time. 

While society as a whole, as well as individuals in particular, have gained a lot from this standardization of time, an essential element has been lost, that of the subjective perception of time. Each person has indeed experienced the very variable perception of the passage of the same unit of time.

At individual level, the perception of the passage of time as well as its duration, varies from one person to another and also at different moments of the life of the same person  \cite{o1, o2, o3, o4}. The passage of time sometimes seems to slow down, speed up or repeat itself \cite{p1, p2, p3}.

Depending on the situation, an hour can seem like five minutes and sometimes like several hours. And that is the question addressed in this paper. Is it feasible to objectively measure the subjective perception of the passage of time? To provide a tentative answer to this question I build a simple model of subjective time passage using a physicist-type approach  \cite{first}.

My mere hypothesis is, given an objective unit of time like the year, to introduce a subjective unit of time for each person, which is inversely proportional to the number of objective units already experienced by that person.  The counting of this number starts from their own birth till the present time. Therefore, the subjective unit of time gets smaller after each new lived objective unit. 

This subjective unit of time allows introducing clock to measure the passage of time as the addition of subjective units in parallel to the addition of the objective units. A future horizon, as well as an origin of the past, are then defined, along with a speed of subjective time flow.

In a second step, I consider ritualized socializations, which span the lifetime of every person, similarly to the birth event, creating additional counting of the passage of time. The model is then extended by stacking individual clocks, each being linked to a social ritual.

The associated equations show that the price for the first ritualized socialization is to exit eternity in terms of a future to be lived with the simultaneous reward of experiencing a moment of infinity. The results recover common feelings about the passage of time over a lifetime. In particular, the fact that time passes more quickly when aging is obtained.

The rest of the paper is organized as follows: The definition of a subjective unit of time associated to an objective unit of time is presented in Section 2? As a result of a subjective unit of time, three subjective features are obtained in Section 3 with the future horizon, the past horizon and the speed of passage of time. Section 4 introduces the setting of ritualized socializations, which creates additional subjective clocks. The outcome of having stacking of clocks is shown in Section five. Concluding remarks are given in last Section.

\section{At the Beginning is an Observation}

The time clock is omnipresent in everyone's life and understood by all. It regulates all moments of life at both levels of the individual and the collective. The measured time is deployed mainly along seconds, minutes, hours, days, weeks, months, years, decades, centuries, millennium. On this basis, I denote $U_o$ any one of these various units. In the following I select the year as this objective unit.

While a year is perfectly well defined for everyone, its duration is perceived differently on an individual subjective level from person to another and at different times in the same person's life. To introduce an element common that is common to all, but but specific to each individual, I consider an individual's lived experience in terms of their total number of years lived.

Thus, for a one-year-old child, a year will be their whole life, while for a six-month-old infant, it is completely undefined, and for a hundred-year-old adult, it is only one hundredth of their life. Thus, every individual can define objectively a subjective unit of time by apprehending their full life as a unitary whole in terms of the chosen objective unit of time.

Given $U_o$, I denote $T_0$ the associated origin of time counting, which is both universal and independent of individuals. For years, this origin has been set at different historical moments by each civilization and religion. Here, I use the origin denoted the Common Era (CE).

Along this objective linear time scale associated with the pair $(U_o, T_0)$, the birth of a given individual is located at some year $T_1$. Thus, at a time $T$, the person has lived $n$ years where $n=T-T_1$. For this given individual, their subjective perception of the chosen objective unit, here the year, is thus the quantity defined as, 
\begin{equation}
U_{s1}=\frac{1}{T-T_1} U_o ,
\label{us1} 
\end{equation}
where $T$ and $T_1$ are measured in units of $U_o$ from the same origin $T_0$. This definition of $U_{s1}$ makes explicit its dependence on $T$, the moment when the evaluation is made. 

Therefore, the unit $U_{s1}$  varies from one individual to another via $T_1$ and also for the same individual, during the objective passage of time via $T$. From Eq.(\ref{us1}) we recover that for $T=T_1+1$, $U_{s1}=U_o$ and for  $T=T_1+10$, $U_{s1}=0.1 \times U_o$.

\begin{figure}[t]
\includegraphics[width=1\textwidth]{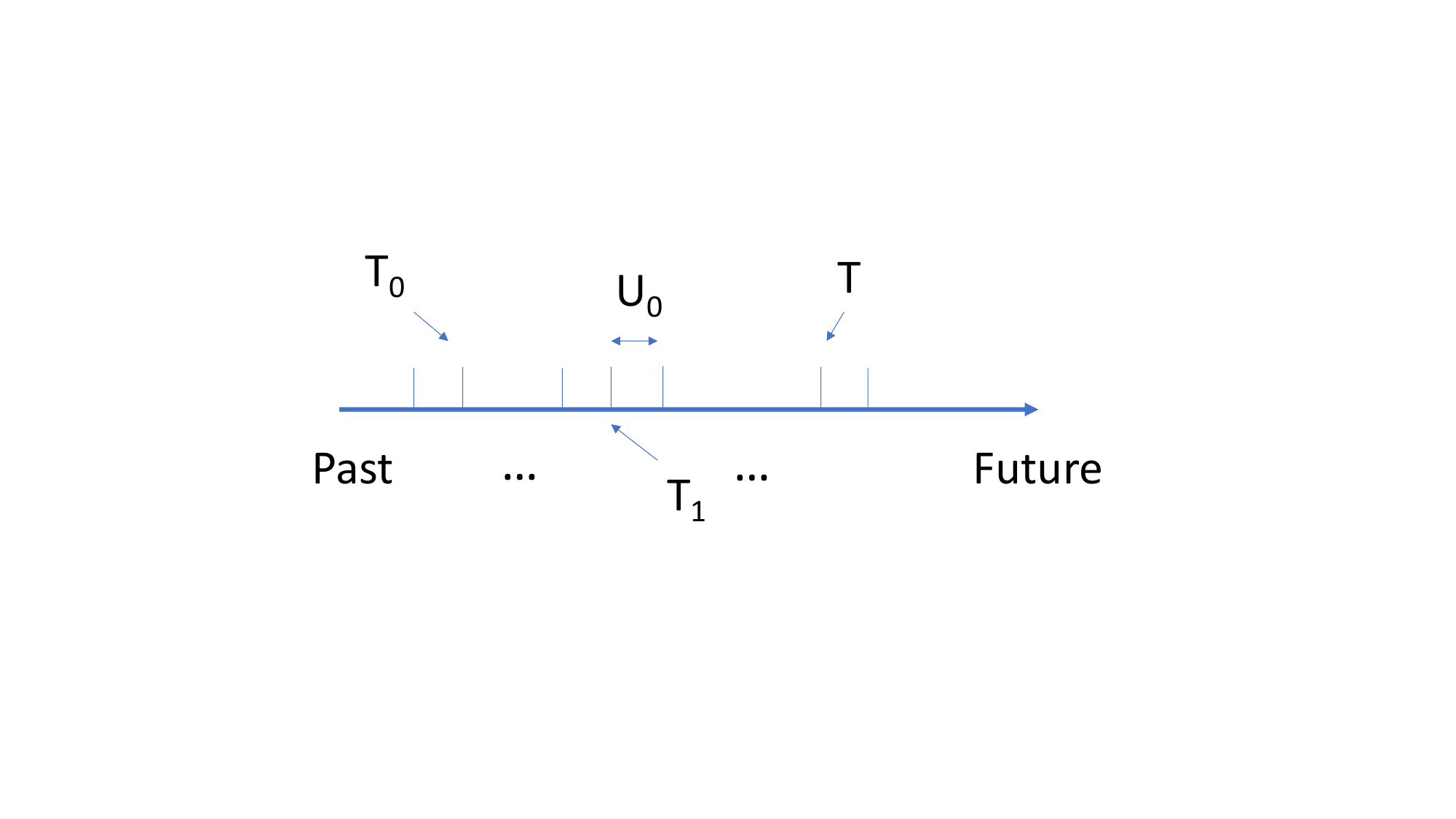}
\caption{Schematic representation of the linear objective time: $T_0$ is the collective origin of time counting, $T_1$ the birthdate of a given person, $T$ the current time and $U_o$ an objective unite of time.}
\label{f1}
\end{figure} 

Moreover, it is interesting to note that for the entire period $T_1\leq T<T_1+1$, $T=T_1$, which gives the surprising result of an infinite unit $U_{s1}\rightarrow + \infty$. This result means that during the passage of time preceding the accomplishment of the first objective unit of time, the subjective unit is perceived as infinite. During a child first year of life, their related perception of time is infinity. It is the discretization of time  \cite{dis}, in terms of objective units, that produces this effect of infinite time feeling.

\section{Three subjective features of a subjective perception of time}

The objective definition of the subjective unit of time $U_{s1}$ implies that when objective time passes, the related subjective unit decreases. For a given person, the perception of the year is thus increasingly shorter as more objective years have been lived. This effect allows defining several characteristic quantities of the global subjective perception of the passage of time for each person.

\subsection{Future horizon}
The first characteristic quantity is the future horizon $H_{f1}$ perceived by a given person who projects into the future, evaluating  the time remaining to reach a specific deadline. To calculate the subjective evaluation of this time I assume that at present time $T$, when an individual born at $T_1$ looks ahead at the time horizon encompassing the next $m$ years, they automatically add up all the related subjective years from this moment $T$ till $T+m$.  At time $T$ the future horizon for the next $m$ years thus writes,
\begin{equation}
H_{f1}=\sum_{t=1}^{m} \frac{1}{T-T_1+t} U_o ,
\label{hf1} 
\end{equation}
which can be rewritten as,
\begin{equation}
H_{f1}=\Big[\sum_{t=1}^{T-T_1+m} \frac{1}{t} -\sum_{t=1}^{T-T_1} \frac{1}{t} \Big]U_o ,
\label{hf2} 
\end{equation}
which is the difference between two harmonic numbers $H_{T-T_1+m}$ and $H_{T-T_1}$. Using their respective asymptotic limits, I get,
\begin{equation}
H_{f1}=\log \Big(\frac{T-T_1+m}{T-T_1} \Big)U_o .
\label{hf3} 
\end{equation}

\begin{figure}[t]
\includegraphics[width=0.9\textwidth]{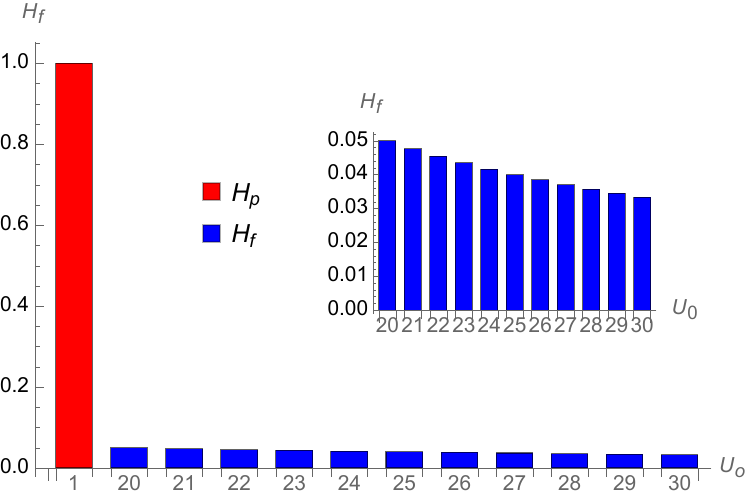}
\caption{The series of successive single year contributions (in blue) for a twenty years old person looking ten years ahead. The first year objective year (in red) is included to show the steep shrinking of the subsequent subjective years with at twenty year old the perception of only $0.05 U_o$. The inset exhibits the drastically shortening of the subjective unit of time towards zero.}
\label{f2}
\end{figure}

From Eq.(\ref{hf3}) the variation of $H_{f1}$ as a function of $m$ is found to be paradoxical. Looking for instance at a case
of a person born at $T_1=2004$ looking at time $T=2024$ at 10 years ahead, i.e., at time $T=2024$, their objective age is $20$ years and their related perceived future horizon equals to,
\begin{eqnarray}
\label{hf4}
H_{f1}&=&\{  \frac{1}{20}+\frac{1}{21}+\frac{1}{22}+\dots +\frac{1}{30} \} U_o   \\ 
&=& \{0.05+0.048+0.045\dots +0.033 \}  U_o \nonumber \\
&\approx & 0.4472 U_o  ,\nonumber
\end{eqnarray}
which is lower than one objective year. Fig.(\ref{f2}) shows the series of successive single contributions from each additional future year. The first year objective year is included to show the steep drop of the subsequent subjective years with at twenty year old the perception of only $0.05 U_o$. The inset in Fig.(\ref{f2}) exhibits the drastically shortening of the subjective unit of time towards zero.

However, for the same person looking at the future without a specific date, Eq.(\ref{hf4}) becomes,
\begin{equation}
H_{f1}=\Big[\frac{1}{20}+\frac{1}{21}+\frac{1}{22}+\dots +\frac{1}{30}+\dots +\frac{1}{\infty}  \Big ]  U_o   ,
\label{hf5} 
\end{equation}
which includes an infinite number of years since they have no a priori reason to set an upper limit on their life in terms of longevity. People do not include a date for their own death. 

Accordingly, although each new term $\frac{1}{T-T_1+t}$  added becomes smaller and smaller since $(T-T_1+t)$ becomes larger and larger,  the number of terms added grows faster than $\frac{1}{T-T_1+t}$ decreases. As a result, the associated future horizon tends to infinity ($H_{f1}\rightarrow + \infty$), despite having every additional subjective year tending to zero. Indeed, this infinite sum of infinitely small terms is itself infinite, I denote this limit as a ``soft infinity." The future horizon is therefore limitless. It is not visible.

\subsection{Past horizon}

The second characteristic quantity is the symmetrical quantity of $H_{f1}$ in time. It is obtained when the individual turns to the past looking at their horizon of the past $H_{p1}$. As with $H_{f1}$ at the present time $T$, they add up all their subjective years already lived since their birth $T_1$, but now in reverse, i.e., backwards in time. And unlike the future, where there is no limit in principle, for the past, there is one, the birth. The past horizon writes,
\begin{equation}
H_{p1}=\sum_{t=0}^{T-T_1} \frac{1}{T-T_1-t} U_o ,
\label{hp1} 
\end{equation}
which gives for the individual born in $2004$ at time $T=2024$,
\begin{eqnarray}
\label{hp1bis} 
H_{p1}&=&\{\frac{1}{20}+\frac{1}{19}+\frac{1}{18}+\dots +\frac{1}{1}+\frac{1}{0} \}  U_o  \\ 
&=& \{0.05+0.053+0.056\dots +1+\infty \}  U_o \nonumber \\
&\approx & \{3.598+\infty \}  U_o \nonumber
\end{eqnarray}

\begin{figure}[t]
\includegraphics[width=0.9\textwidth]{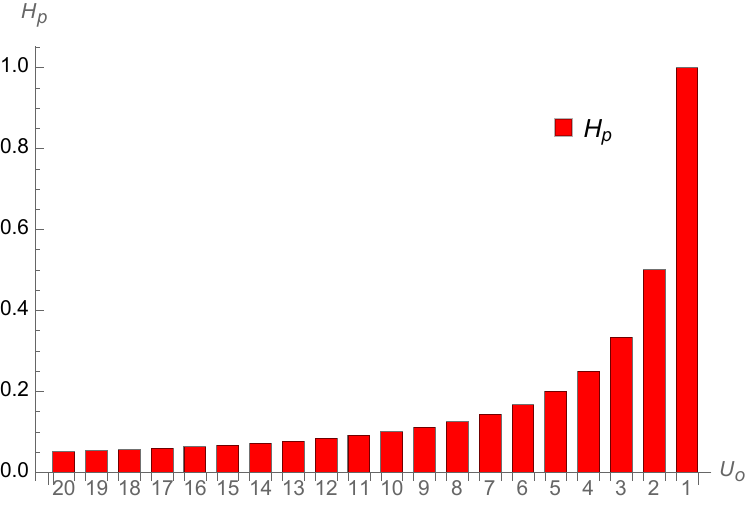}
\caption{The series of single contributions from each past year without the infinite first one. Contrary to the case of the future horizon, for the past horizon, each term gets larger till one just before infinity.}
\label{f3}
\end{figure}

Fig.(\ref{f3}) shows the series of single contributions from each l past year without the infinite first one. Contrary to the case of the future horizon, for the past horizon, each term gets larger till one just before infinity.

\begin{figure}[t]
\includegraphics[width=0.9\textwidth]{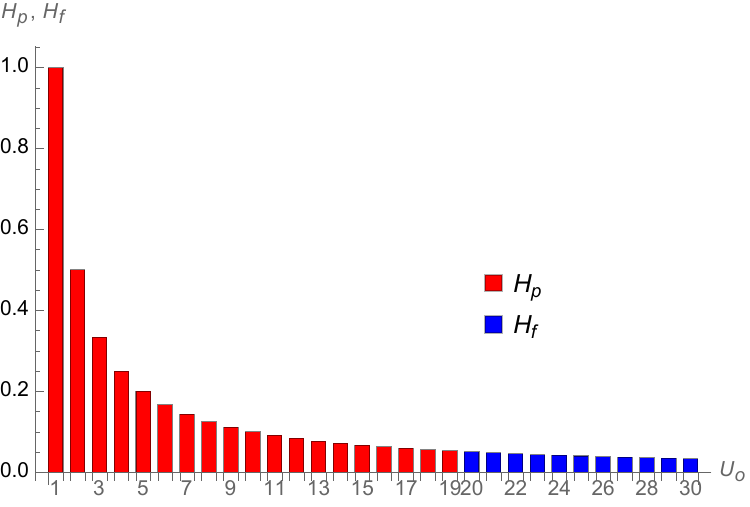}
\caption{Both the past Horizon (in red) without the infinite initial term and the future horizon (in blue) for a finite time projection at ten years ahead.}
\label{f4}
\end{figure}

Although this time the number of terms in the sum is finite, the result is still infinite with $H_{p1}\rightarrow +\infty$. But here, the infinity is of a different nature than for the future horizon. Indeed, it is only the last term of the sum $\frac{1}{0}$ at $t=T_1$ (birth) that corresponds to the first experimentation of the unit of time, which creates the limit to infinity. I denote it a "hard infinity." Fig.(\ref{f4}) includes both the past Horizon (without the infinite initial term) and the future horizon for a finite time projection at ten years ahead.

\subsection{The subjective speed of time}

\begin{figure}
\includegraphics[width=0.9\textwidth]{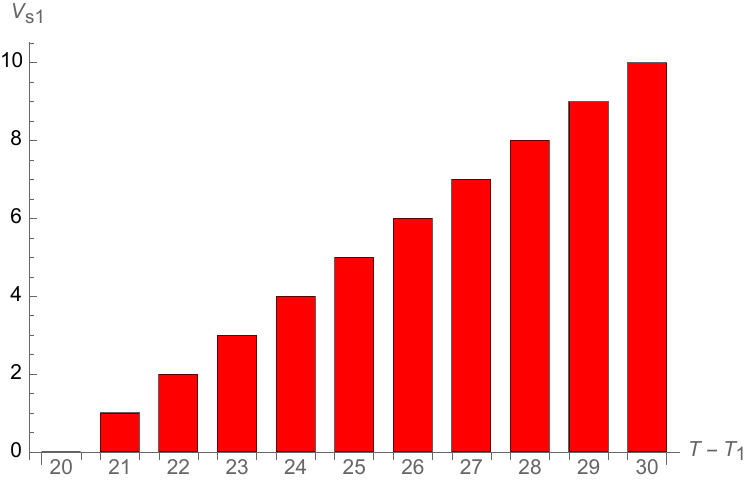}
\caption{The increase in speed with the passing years from age twenty till age thirty. The slope is linear with the number of lived years as seen }
\label{f5}
\end{figure}

At this stage, I define at time $T$ a third characteristic quantity with a subjective flow rate of time $V_{s1}$ by noting that the objective unit  $U_o$ corresponds to the subjective flow rate $U_o$ multiplied by the subjective unit $U_{s1}$, which yields,
\begin{equation}
U_o=V_{s1} U_{s1} ,
\label{v1} 
\end{equation}
which gives using Eq.(\ref{us1}),
\begin{equation}
V_{s1}=n=(T-T_1) ,
\label{v1bis} 
\end{equation}
where $n$ is the number of objective units of time already experienced. The increase is linear with the number of lived years as seen in Fig.(\ref{f5}).

In other words, the subjective speed of time is just the inverse of the subjective unit. And thus, the more one has lived, the faster time passes (the speed increases), while conversely the subjective duration of the subjective unit becomes shorter. 

Intuitively, this result can be understood saying that since the unit of time is shortening, one must go faster to cover the same objective time distance. It is like with a car where the distance traveled would correspond to objective time. The duration of the journey would represent subjective time. Therefore, the shorter the latter, the greater the speed must be to cover the same distance of objective time. 

For the individual born in 2004, in 2024, their flow rate is 20 objective years per subjective year. In other words, to experience the passage of one objective year, this person must experience the passage of twenty related subjective years, ie, to move over twenty objective years during the perception of one subjective year.

\subsection{Main preliminary findings}

At this stage, my hypothesis of a subjective unit of time given by Eq.(\ref{us1}) has led to the following three main consequences with respect to the subjective perception of time:

\begin{enumerate}

\item The associated future horizon is imperceptible, with the sensation of an endless life to be lived (soft infinity).

\item In parallel, the past horizon is not localized with the sensation of a start from an infinite time ago (hard infinity).

\item With the passage of time, a narrowing of the subjective unit of time produces a related acceleration of its flow. The older a person gets, the faster they experience the passage of time.

\end{enumerate}

In summary, the subjective perception of the passage of time by every person would therefore be infinite, without neither a beginning nor an end, and with a simultaneous increase of the speed of passage of time.

\section{From birth to first ritual socialization}

However, it is remarkable to note that all human societies have integrated a series of additional ``social births" in addition to the birth. These social births are set via implementing social rituals like communion in most religions, graduation during studies, military service in some countries, weddings, birth of children, divorce.
 
The ritualized social events are always in small numbers and concentrated mostly in the first thirty years of life. They shapes the lifetime of each member of a collective community with a series of dates $T_2, T_3, T_4, \dots$,  which punctuate the shape of the passage of time in everyone lifetime. 

New counting clocks are thus ``naturally" added to the $T_1$ initial birth clock, which marked the beginning of counting of objective lived time units. To incorporate these new clocks within the framework of the model, I consider the introduction of the first significant event at a date $T_2$. Then, I assume that a second clock is created in addition and similarly to that of birth.

Applying the previous mechanism of constructing the subjective unit $U_{s1}$ instead of $U_o$, I define a new subjective unit of time, which  at time $T$ writes,
\begin{equation}
U_{s2}= \frac{1}{T-T_2} U_{s1} ,
\label{uu1} 
\end{equation}
or using Eq.(\ref{us1}),
\begin{equation}
U_{s2}=\frac{1}{(T-T_1)(T-T_2)}U_o .
\label{uu2} 
\end{equation}

As before, during the period $T_2\leq T<T_2+1$, $T=T_2$, which recovers the effect of a hard infinity with ($U_{s2}\rightarrow \infty$). Thus, the reward for the first ``socialization" through the ritual, for example, of marriage, is to recreate for the individual the initial sensation of infinity felt during their first year of existence just after birth. As a result, the individual is quite happy with the initiation of a second clock, in addition to the first. The subsequent effect is the creation of  a sequential memory of different subjective times.

\subsection{Past horizon}

With respect to the past horizon, at time $T$ two different backward sums have now to be considered. The first corresponds to the current unit $U_{s2}$, from $t=T$ back to $t=T_2$, and the second, with the subjective unit $U_{s1}$, from $t=T_2$ back to $t=T_1$. Thus,
\begin{equation}
H_{p2}= \Big [\sum_{t=T_2}^{T} \frac{1}{(t-T_1)(t-T_2)}  + \sum_{t=T_1}^{T_2} \frac{1}{t-T_1} \Big ]  U_o .
\label{hp2} 
\end{equation}

The sum is doubly infinite due to the two terms at $t=T_1$ and $t=T_2$. The sensation of an infinite origin is thus preserved, but it becomes blurry with the feeling that the origin is even further away due to the addition of two different infinities.

For the case of the individual born in $2004$, if married at $T_2=2034$, their origin of the past at $T= 2044$ would be,

\begin{eqnarray}
\label{hp2bis} 
H_{p2}&=& \Big [ \frac{1}{10 \times 40}+\frac{1}{9 \times 39}+\dots +\frac{1}{1 \times 31}+\frac{1}{0 \times 30}\\ \nonumber
&+& \frac{1}{30}+\frac{1}{29}+\frac{1}{28}+\dots \frac{1}{1}+\frac{1}{0} \Big ]  U_o , \\
&=&  [0.0025+0.0028 +\dots +0.032 + \infty    \nonumber \\
&+&  0.033+0.035 +0.036+\dots +1+ \infty  ]  U_o  \nonumber \\
&=&  [ 3.995+0.088+ 2 \times +\infty  ]  U_o . \nonumber 
\end{eqnarray}

The outcome is thus similar to the case prior to the first ritual socialization as seen comparing with Eq.(\ref{hp1bis}).

\subsection{Future horizon}

However, while the introduction of a second clock did not change qualitatively the perception of past horizon, a qualitative change does occur for the future horizon $H_{f1}$, which writes,
\begin{equation}
H_{f2}=\sum_{t=1}^{m} \frac{1}{(T-T_1+t)(T-T_2+t)} U_o .
\label{hf1} 
\end{equation}

Noticing that although the sum is over an infinite series of decreasing terms as before, now the successive terms vary as $\frac{1}{t^2}$ instead of $\frac{1}{t}$, which in turn has a drastic impact on the limit when the horizon is set at infinity ($m\rightarrow +\infty)$.

Indeed, it happens that $\frac{1}{t^2}$ tends faster to zero than the number of added terms, and thus the sum $H_{f2}$ no longer diverges towards infinity as $H_{f1}$ does. The future horizon becomes bounded, i.e., its numerical value is finite. The asymptotic difference appears clearly in the continuous version of the time. In this case, the horizon is obtained using the integrals of the subjective units $\frac{1}{t^2}$ and $\frac{1}{t}$, which yield respectively $-\frac{1}{t}$ and $\log t$ whose limits are $0$ and $+\infty$ for $t \rightarrow +\infty$.

This qualitative change in the perception of time implies that the future horizon suddenly becomes visible. Projection in the ``infinite" future is now at a finite distance from the present. The first ritualized socialization turns ``eternal "humans to ``mortal" humans with respect to the perception of their ending future.

Pushing the logic of reasoning, I could hypothesize that refusing any ritual socialization would lead to a subjective eternal life. But at the same time, what is a life outside of any collective socialization? May be this result could be linked to autism. However, while this connexion deserves further exploration it is beyond my expertise.

\subsection{Time acceleration} 

The updated subjective unit of time being $U_{s2}$ instead of $U_{s1}$ the associated subjective speed of the passage of time becomes,
\begin{equation}
V_{s2}=(T-T_1)(T-T_2) .
\label{v2} 
\end{equation}

After the ritual socialization, which has introduced a second subjective clock, the passage of time gets busted to pass by more quickly as a function of $T^2$ instead of $T$.

\section{The stacking of clocks}

The first clock-like socialization has created the condition to live again the feeling of eternity during the completion of the following year. However, the associated cost has been an increase of the passage of time and a shrinking of the subjective life expectancy. 

Empowered by this paradoxical life experience, people have been adding more social ritual to create additional clocks for the subjective passage of time. I could formulate the process as If to die, might as well increase the moments of eternity.

Accordingly, all organized human groups have built a series of rituals which are quasi-mandatory to go through. The net result is a stacking of several subjective clocks. The number $L$ of ritual socializations and the associated clocks is a function of the various cultural communities and may vary from one person to another within a given community.

Each new clock is set at a different time yielding a series of initial times $T_2,T_3 \dots,T_L$, which results at a subjective unit of time written as,
\begin{equation}
U_{sL}= \frac{1}{\prod_{i=1}^{L} (t-T_i) }U_o .
\label{uL} 
\end{equation}
where $T_1$ the birthdate.

The subjective unit of time is now a function of $ \frac{1}{T^L}$, i.e., with a convergence to zero extremely rapid. The larger $L$, the more abrupt this fall in the subjective duration. 

As a consequence, the future horizon gets rather close with,
\begin{equation}
H_{sL}=\sum_{t=1}^{m}\frac{1}{\prod_{i=1}^{L} (T-T_i+t) } U_o  , 
\label{hfL} 
\end{equation}
a distance that becomes almost zero, creating the feeling that life is over.

In parallel, the past horizon keeps getting more blurry with the addition of several hard infinities as seen with,
\begin{eqnarray}
\label{hpL} 
H_{pL}&=& \Big [ \sum_{t=T_L}^{T}\frac{1}{\prod_{i=1}^{L} (T-T_i+t) }+\dots \\
&+&\sum_{t=T_2}^{T} \frac{1}{(t-T_1)(t-T_2)}  + \sum_{t=T_1}^{T_2} \frac{1}{t-T_1} \Big ]  U_o .\nonumber
\end{eqnarray}

Along the changes of subjective unit of time, future horizon and origin of the past, the speed of the passage of time becomes,
\begin{equation}
V_{sL}=\prod_{i=1}^{L} (T-T_i) ,
\label{vL} 
\end{equation}
indicating a high value, which increase like $T^L$ where $L$ is the number of clocks.

\section{Damping the the proportionality effect}

At this stage, the main consequences of the hypothesis of the inverse proportionality of  Eq. (\ref{us1}) have been obtained. The associated trends are qualitatively sound but may be quantitively inappropriate. In particular, the magnitude of $U_s$ seems to shrink too fast with the increasing of the number $n$ of years lived.  

\begin{figure}[t]
\includegraphics[width=1\textwidth]{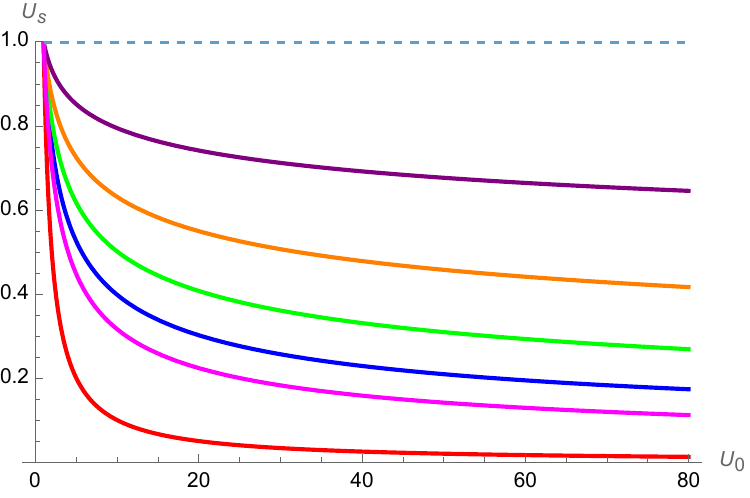}
\caption{Variation of the subjective unit of time $U_{sa}$ as a function of the number of past objective unite of time for the series of exponents $a=1, 0.5, 04, 0.3, 0.2, 0.1, 0$, with $a=1$ (in red) for the lowest curve and $a=0$  for the upper horizontal curve (in dots).}
\label{f6}
\end{figure}

While the decrease of $U_s$ with respect to of $U_o$ is sound for the first tow or three years, a damping may be in order to make the definition more realistic. One possible path is to introduce a power law as,
\begin{equation}
U_{sa}=\frac{1}{(T-T_1)^a} ,
\label{us2} 
\end{equation}
with $0< a < 1$. The power law reduces substantially the decreasing of $U_s$ with $n$ as seen in Fig. (\ref{f6}). The value of the exponent $a$ could be evaluated via a fitting to relevant data. It could a feature, which is either universal, i.e., the same for everyone or specific to each person.

\begin{figure}[t]
\includegraphics[width=1\textwidth]{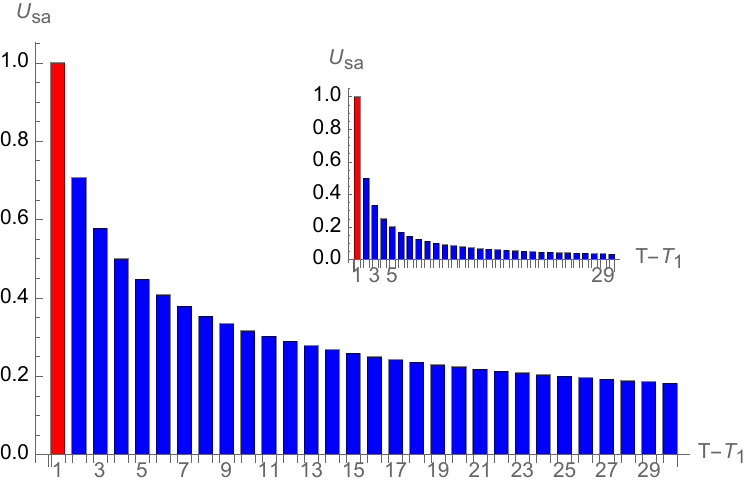}
\caption{The variations of $U_{sa}$ for the first thirty years of existence of a person in the case $a=0.5$, which means a square root dependence. The inset shows the  variation of of $U_{sa}$ for $a=1$.}
\label{f7}
\end{figure}

To illustrate the damping effect, Fig. (\ref{f7}) shows the variations of $U_{sa}$ for the first thirty years of existence of a person in the case $a=0.5$, which means a square root dependence. The inset shows the  variation of of $U_{sa}$ for $a=1$.

The qualitative behavior of the future horizon, the past horizon and the speed of time are preserved using a power law with only the quantitative values being modified.

 \section{Conclusion}

I have presented a stylized model, which, although very simple, is able to yield a coherent approach to describe the subjective passage of time at an individual level. Given an objective unit of time, my main hypothesis introduces an actual mirror subjective unit of time as a function of the number of associated objective units of time already experienced by a person. This subjective unit of time varies as $1/n$ where $n$ is the number of objective units the person has already experienced. Moreover, this definition of a  dynamics subjective unit of time allows to define a future horizon, the past horizon and the speed of time, with quantitative formulas. 

Following above construction, I reproduced the same mechanism introducing a first ritualized socialization, which adds a ``second birth" and then a second counting of the passage of time. The associated subjective unit of time now varies as $1/n^2$ instead of $1/n$. Along this path, several additional subjective clocks associated with more ritualized socializations, are stacked to shape a subjective unit of time, which has a series of singularities.

The associated equations showed that the price for the first ritualized socialization is to exit the subjective feeling of eternity in terms of a future to be lived with the simultaneous reward of experiencing another moment of infinity similar to the birth. The results recover common feelings about the passage of time over a lifetime, with particular, the fact that time passes more quickly when aging is obtained.

Noticing that the amplitudes of the subjective units of time sound too small, I rescaled the number of years a person has experienced using a power law. The qualitative results are conserved with a simultaneous damping of the shrinking of time. The value of the exponent could be determined using some data about the actual feeling of the passage of time reported by people.

\end{document}